# New measurements and analysis of the β Cephei star V909 Cassiopeiae

David R. S. Boyd [1], Robert A. Koff [2]
[1] West Challow Observatory, Variable Star Section, British Astronomical Association [davidboyd@orion.me.uk]
[2] Antelope Hills Observatory [bob@antelopehillsobservatory.org]

**Abstract**

V909 Cas is a little-studied example of a β Cep pulsating variable located in the OB association Cas OB8 in the Perseus spiral arm of the Milky Way. Photometric observations in 2016-7 provided 30 new times of pulsation extrema and enabled its mean pulsation period to be determined as 0.2067798(1) d. From spectroscopic observations we determined its interstellar extinction and absolute magnitude and luminosity, and located it with other β Cep stars in the OB instability region of the H-R diagram.

**β Cephei variables**

β Cep stars are the brightest recognised class of variable stars in the sky, although many other highly luminous stars are intrinsically variable. They are pulsating stars whose brightness varies with periods of 0.1 to 0.3 days and with amplitudes of 0.1 magnitudes or less. Most have spectral types in the range B0 to B2 and luminosity classes III to V (1). They lie within the OB instability strip located on the main sequence in the Hertzsprung-Russell diagram. The pulsation of β Cep stars is due to the *kappa* mechanism acting on iron-group atoms deep within the star (2). In some β Cep stars the pulsation amplitude varies because of beating between unresolved pulsation frequencies. Stankov & Handler (3) review current knowledge about β Cep stars and list 93 confirmed and 77 suspected instances. In the General Catalogue of Variable Stars (4) β Cep stars are designated as BCEP. The AAVSO Variable Star Index (VSX) (5) currently lists 255 BCEP stars, three-quarters of which are south of the celestial equator. This is consistent with stars closer to the galactic centre having the higher metal abundancies conducive to the β Cep instability mechanism (6). OB associations comprise young stars within galactic Population I and are concentrated towards the galactic plane.

**V909 Cassiopeiae**

In the course of observations of the X-ray source RX J0136.7+6125, Robb et al. (7) serendipitously discovered that the nearby star BD+60 282 (GSC 04031-00631) was a β Cep type pulsating star. It was subsequently given the designation V909 Cas. They recorded three series of R filtered photometry of the star between December 1996 and January 2000 during which they measured 18 times of maximum of the light curve and 11 times of minimum. From these they derived three independent values for its pulsation period. Calculating a weighted mean of these periods gives 0.2066(2) days where the figure in parentheses is the uncertainty in the final digit. They also noted that the pulsation amplitude varied between 0.01 and 0.05 magnitudes, and it appears from their Figure 3 that the times of maximum varied slightly with respect to the pulsation phase. V909 Cas is one of the relatively few β Cep stars for which VSX does not currently list a pulsation period. We decided to investigate its current behaviour.

**Photometric observations**

The coordinates of V909 Cas are RA 01h 36m 38.55s, Dec +61° 25' 54.0". We obtained images of this field using B, V and Rc filters on 11 nights between December 2016 and December 2017 using 0.35m (DB) and 0.25m (RK) Schmidt-Cassegrain telescopes and CCD cameras. Table 1 gives a log of these observations. By obtaining observations from two widely separated longitudes (-4° W and –104° W) we mitigated the potential problem of ±1 day alias solutions when analysing our results.

Comparison stars were chosen close to the variable, excluding RX J0136.7+6125 which was found to be variable by Robb et al. (7), and their magnitudes obtained from the APASS database (8). Images were dark-subtracted, flat-fielded and the magnitude of the variable established by differential aperture photometry with respect to comparison stars. Where two filters were used alternately during an observing run, magnitudes were transformed to the Johnson-Cousins standard system. Light curves were constructed for each observing run and each filter. Figure 1 shows V-band light curves for 2016 December 28, 2017 January 2 and January 18 phased on the pulsation ephemeris in eqn (1) derived below. These show variation of the pulsation amplitude over a period of 3 weeks and close inspection also shows that the light curves drift slightly in phase over time. We measured a mean V magnitude $m_v$ = 10.63(1) and colour indices (B-V) = 0.45(1) and (V-R) = 0.27(1). These colour indices remained constant through the pulsation cycle.

Times of maximum and minimum of the pulsation cycle were obtained by fitting a quadratic function to the extrema of each light curve. These times are listed in Table 2. The uncertainties in these times are larger when the pulsation amplitude is small because the extrema are then less clearly defined. In runs which included both a maximum and a minimum it was possible to determine the pulsation amplitude. These are listed in Table 3 and confirm the large variation found by Robb et al. (7), although Figure 1 shows that we find the variation to be approximately symmetrical about the mean magnitude level rather than having a constant minimum magnitude as they found.

**Pulsation ephemeris**

The pulsation behaviour of stars like V909 Cas is complex with many periods involved but these may combine to yield a single effective period, although as noted above this may also result in the phase of the pulsation light curve drifting about the effective period. As we have no way of determining the number of pulsation cycles which have occurred between 2000 and 2016, we adopted a heuristic approach to determining the pulsation ephemeris using the times of 29 extrema measured between 1996 and 2000 and 30 extrema obtained in 2016-2017. The pulsation period was varied incrementally in steps of 0.000001 d between 0.2056 d and 0.2076 d, a range equivalent to ±5 standard deviations from the value found by Robb et al. (7). At each step the pulsation period was used to predict the times of extrema of the pulsation cycle relative to a well-measured maximum on 2017 January 13. We assume maxima occur at phase 0 and minima at phase 0.5. The times of these predicted extrema were compared to times of the measured extrema to see if they were within a quarter phase of correctly predicting whether the extremum was a maximum or minimum. This latitude allows for some drift in phase of the light curve from cycle to cycle.

Only six solutions were found in which all 59 extrema were predicted correctly. The root mean square (rms) residual between the predicted and observed times of extrema was calculated for each solution. Of these six solutions, there was one with a distinctly lower rms residual than the others. This corresponded to a mean pulsation period of 0.2067798(1) d. To test the robustness of this result, the search was repeated predicting the times of extrema relative to the measured times of several different recent maxima. The search was also repeated using only the recent V-band data and only the recent R-band data. In every case the same preferred period was found.

A linear ephemeris for maxima of the pulsation cycle of V909 Cas is:

$$\text{HJD (maximum)} = 2457767.351(1) + 0.2067798(1) * E \qquad (1)$$

To investigate this further, we analysed all our V-band data for 2016 and 2017 by the phase dispersion minimization (PDM) technique in Peranso (9). This method of analysis uses the whole light curve, not just the times of extrema, but has to contend with the continually varying pulsation amplitude and possibly a lack of strict phase coherence. This analysis produced a broad minimum with multiple "valleys", the deepest of which was at 0.20678(3) d consistent with the above ephemeris.

The residuals of the measured times of extrema with respect to the linear ephemeris in eqn (1) are also included in Table 2. Given that the extrema do not follow a strictly regular clock as demonstrated by the drift in phase noted above, it is not surprising that the residuals are on average larger than the uncertainties in the measured times of extrema.

**Spectroscope observations**

We obtained low resolution (5Å) spectra of V909 Cas with a LISA spectrograph from Shelyak Instruments (11) on a 0.28m SCT. The spectra were bias and dark subtracted, flat fielded, wavelength calibrated using the spectrograph's internal Ar-Ne lamp and corrected for instrumental and atmospheric response using spectra of a star of known spectral type at the same air mass as V909 Cas recorded immediately prior to the spectra of V909 Cas.

McCluskey et al. (10) give the spectral type of BD+60 282 (V909 Cas) as B1III. Luminosity class III indicates it is a giant star. Our spectra appeared to be of a much later spectra type indicating that light from the star had been heavily reddened as it passed through the interstellar medium. Assuming Rv, the ratio of absolute to relative interstellar extinction in the V-band, is 3.1, Cardelli et al. (12) give a parameterisation of the wavelength dependence of absolute extinction A(λ) relative to visual extinction A(V). From this parameterisation the wavelength dependence of relative extinction or reddening, E(λ -V), can be determined for a given value of the colour excess E(B-V).

We investigated the value of E(B-V) which would deredden the observed spectrum to best match spectral type B1III and found the closest match with E(B-V) = 0.72(1). Figure 2 shows the spectrum of V909 Cas we observed on 2016 December 28, the same spectrum dereddened with E(B-V) = 0.72 and the Pickles Stellar Spectral Flux Library (13) spectrum for spectral type B1-2III. Together with our measured colour index (B-V) = 0.45 this gives an intrinsic colour index for V909 Cas of $(B-V)_0$ = -0.27. The spectra showed no perceptible variation over the pulsation cycle.

The literature provides a range of values for the effective temperature Teff of stars with spectral type B1III. From Popper (14), Underhill et al. (15), Boehm-Vitense (16) and Fitzgerald (17) we find a mean value of Teff = 26,300±150 K and hence log(Teff) = 4.420(2). This temperature is consistent with a colour index $(B-V)_0$ = -0.27.

**Location of V909 Cas in the Milky Way**

According to Simbad (18) the galactic coordinates of V909 Cas are l = 128.1°, b = -0.97°. This places the star in the galactic plane in the direction of the Perseus spiral arm. OB associations of bright stars are likely to be found in star-forming regions within spiral arms. Garmany & Stencel (19) assign BD+60 282 (V909 Cas) to OB association Cas OB8. Melnik & Dambis (20) give the distance to Cas OB8 as 2.3(1) kpc based on new reductions of Hipparcos data. This is consistent with the current poorly defined distance to V909 Cas of 2.7±1.6 kpc given in Gaia DR1 (21). Figure 3 shows an artist's concept of the position of the Perseus spiral arm relative to the Sun and the Galactic Centre based on infrared images from the Spitzer Space Telescope (22). The inferred location of V909 Cas is marked with a red dot.

**Location of V909 Cas in the H-R diagram**

Assuming Rv = 3.1, a colour excess E(B-V) = 0.72(1) gives interstellar extinction in the V-band, Av = 2.23(3) magnitudes. This is consistent with the average value of extinction in the solar neighbourhood of 0.7 to 1 mag/kpc. Having measured the mean V-band apparent magnitude of V909 Cas as mv = 10.63(1), and knowing d = 2300±100 parsec, we can calculate Mv, the absolute V magnitude of V909 Cas as

$$Mv = mv - Av - 5\log(d/10) = -3.4(1) \qquad (2)$$

Adopting a mean bolometric correction for B1III stars of -2.56(5) from Popper (14) and Flower (23) gives a bolometric magnitude for V909 Cas of Mbol = -5.96(11). Taking the bolometric magnitude of the Sun as 4.75 we can calculate the absolute luminosity L of V909 Cas relative to that of the Sun $L_\odot$ as

$$\log(L/L_\odot) = -0.4(Mbol - 4.75) = 4.28(4) \qquad (3)$$

Figure 3 in Pamyatnykh (2) shows the location of β Cep stars within the OB instability region in the H-R diagram plotted with axes log(L/L$_\odot$) and log(Teff). Using that diagram as our Figure 4, with log(Teff) = 4.420(2) and log(L/L$_\odot$) = 4.28(4), we have marked the position of V909 Cas as a red dot showing that it lies among other β Cep stars in the OB instability region. The size of the dot encompasses the uncertainties in the parameter values. From Figure 5 of Stankov & Handler (3), these parameter values indicate the mass of V909 Cas is approximately 13 solar masses. It is approximately 19,000 times as luminous as the Sun and, from the Stefan-Boltzmann Law which relates luminosity, radius and temperature, its radius is approximately 6.7 solar radii.

**Summary**

Our observations of V909 Cas are consistent with it being a typical β Cep pulsating variable located within OB association Cas OB8 in the Perseus spiral arm of the Milky Way. Using published timings of extrema of the light curve from 1996-2000 and our new measurements in 2016-7, we were able to correctly predict the nature of all recorded extrema with a mean pulsation period of 0.2067798(1) d. We observed short term drifts in phase of the pulsation cycle and a V-band pulsation amplitude which ranged from 0.014 to 0.098 mag. We measured its spectrum and, using its published distance and spectral type of B1III, determined its effective temperature, V-band interstellar extinction, absolute magnitude and absolute luminosity, thus locating the star within the OB instability region in the HR diagram.

**Acknowledgements**

This research has made use of the SIMBAD database, operated at CDS, Strasbourg, France, and the International Variable Star Index (VSX) and AAVSO Photometric All-Sky Survey (APASS), both operated at AAVSO, Cambridge, Massachusetts, USA. This work has made use of data from the European Space Agency (ESA) mission Gaia processed by the Gaia Data Processing and Analysis Consortium (DPAC). We are grateful for the comments of the referees which have assisted us in improving the paper.

| Date | Start of run (JD) | Duration (hr) | No of images | Filters | Observer |
|---|---|---|---|---|---|
| 22 Dec 2016 | 2457745.278 | 3.71 | 888 | V,R | DB |
| 26 Dec 2016 | 2457749.266 | 6.04 | 1786 | V,R | DB |
| 28 Dec 2016 | 2457751.272 | 5.02 | 1268 | V,R | DB |
| 2 Jan 2017 | 2457756.222 | 6.12 | 1980 | V,R | DB |
| 13 Jan 2017 | 2457767.253 | 4.62 | 1262 | V,R | DB |
| 18 Jan 2017 | 2457771.550 | 6.15 | 638 | V | RK |
| 19 Jan 2017 | 2457772.551 | 5.81 | 615 | V | RK |
| 19 Jan 2017 | 2457773.240 | 6.50 | 2272 | V,R | DB |
| 29 Oct 2017 | 2458056.365 | 2.78 | 744 | V,R | DB |
| 12 Nov 2017 | 2458070.245 | 2.71 | 438 | B,V | DB |
| 27 Dec 2017 | 2458115.231 | 4.10 | 814 | B,V | DB |

Table 1: Log of observations.

| Date | Filter | Type of extremum | Time of extremum (HJD) | Uncertainty (d) | Residual (d) |
|---|---|---|---|---|---|
| 26 Dec 2016 | V | Max | 2457749.3059 | 0.0170 | -0.0553 |
| 26 Dec 2016 | R | Max | 2457749.3130 | 0.0177 | -0.0481 |
| 26 Dec 2016 | V | Min | 2457749.4109 | 0.0117 | -0.0537 |
| 26 Dec 2016 | R | Min | 2457749.4234 | 0.0151 | -0.0412 |
| 28 Dec 2016 | V | Min | 2457751.3480 | 0.0104 | 0.0224 |
| 28 Dec 2016 | R | Min | 2457751.3531 | 0.0087 | 0.0275 |
| 28 Dec 2016 | V | Max | 2457751.4356 | 0.0076 | 0.0066 |
| 28 Dec 2016 | R | Max | 2457751.4377 | 0.0073 | 0.0088 |
| 2 Jan 2017 | R | Min | 2457756.2750 | 0.0122 | -0.0133 |
| 2 Jan 2017 | V | Min | 2457756.2850 | 0.0112 | -0.0033 |
| 2 Jan 2017 | V | Max | 2457756.3819 | 0.0068 | -0.0097 |
| 2 Jan 2017 | R | Max | 2457756.3837 | 0.0108 | -0.0080 |
| 13 Jan 2017 | R | Max | 2457767.3427 | 0.0050 | -0.0083 |
| 13 Jan 2017 | V | Max | 2457767.3441 | 0.0046 | -0.0070 |
| 18 Jan 2017 | V | Min | 2457771.5904 | 0.0044 | 0.0004 |
| 18 Jan 2017 | V | Max | 2457771.6971 | 0.0033 | 0.0037 |
| 19 Jan 2017 | V | Min | 2457772.6194 | 0.0038 | -0.0045 |
| 19 Jan 2017 | V | Max | 2457772.7263 | 0.0044 | -0.0010 |
| 19 Jan 2017 | R | Max | 2457773.3503 | 0.0053 | 0.0027 |
| 19 Jan 2017 | V | Max | 2457773.3512 | 0.0044 | 0.0036 |
| 19 Jan 2017 | R | Min | 2457773.4541 | 0.0054 | 0.0031 |
| 19 Jan 2017 | V | Min | 2457773.4573 | 0.0051 | 0.0062 |
| 29 Oct 2017 | R | Max | 2458056.4217 | 0.0066 | -0.0075 |
| 29 Oct 2017 | V | Max | 2458056.4225 | 0.0047 | -0.0068 |
| 12 Nov 2017 | V | Max | 2458070.2943 | 0.0044 | 0.0108 |
| 12 Nov 2017 | B | Max | 2458070.2947 | 0.0041 | 0.0112 |
| 27 Dec 2017 | B | Min | 2458115.2626 | 0.0100 | 0.0045 |
| 27 Dec 2017 | V | Min | 2458115.2626 | 0.0075 | 0.0045 |
| 27 Dec 2017 | V | Max | 2458115.3529 | 0.0068 | -0.0085 |
| 27 Dec 2017 | B | Max | 2458115.3562 | 0.0051 | -0.0052 |

Table 2. New measured times of extrema of the pulsation cycle with uncertainties and their residuals with respect to the linear ephemeris in eqn (1).

| Date | Pulsation amplitude (B-band magnitude) | Pulsation amplitude (V-band magnitude) | Pulsation amplitude (R-band magnitude) |
|---|---|---|---|
| 26 Dec 2016 | | 0.019 | 0.016 |
| 28 Dec 2016 | | 0.015 | 0.015 |
| 2 Jan 2017 | | 0.031 | 0.033 |
| 18 Jan 2017 | | 0.098 | |
| 19 Jan 2017 | | 0.067 | |
| 19 Jan 2017 | | 0.058 | 0.049 |
| 27 Dec 2017 | 0.049 | 0.041 | |

Table 3: Measured pulsation amplitudes.

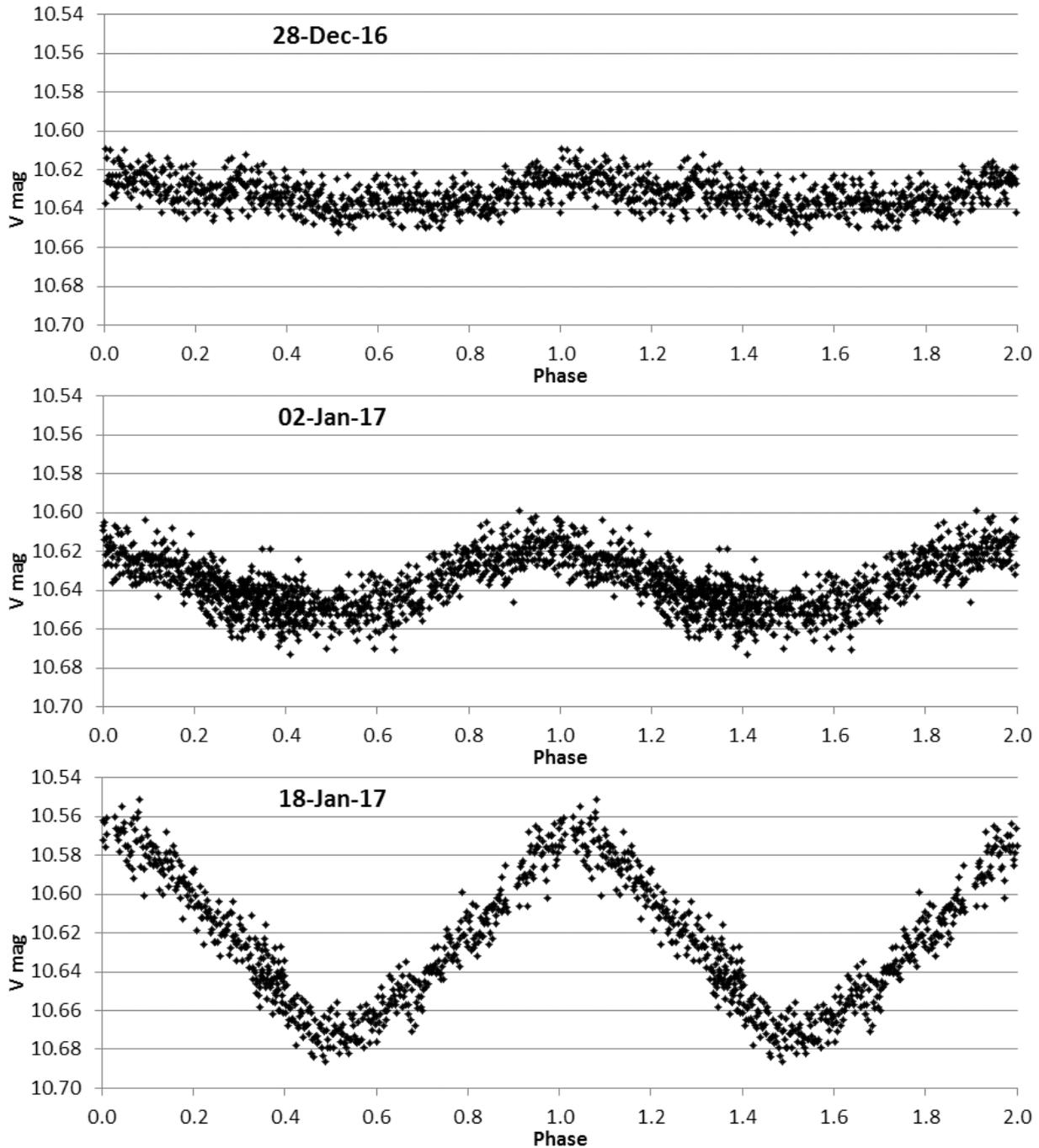

Figure 1. V-band light curves of V909 Cas for 2016 December 28, 2017 January 2 and January 18 phased on the pulsation ephemeris in eqn (1).

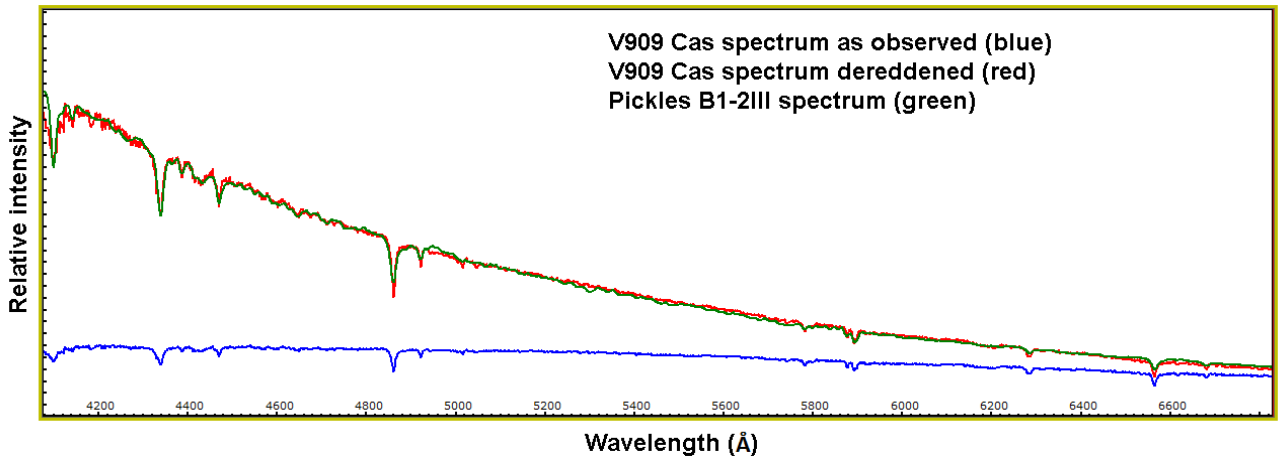

Figure 2. Spectrum of V909 Cas observed on 2016 December 28 (blue), dereddened with colour excess E(B-V) = 0.72 (red), and the Pickles Stellar Spectral Flux Library spectrum for spectral type B1-2III (green).

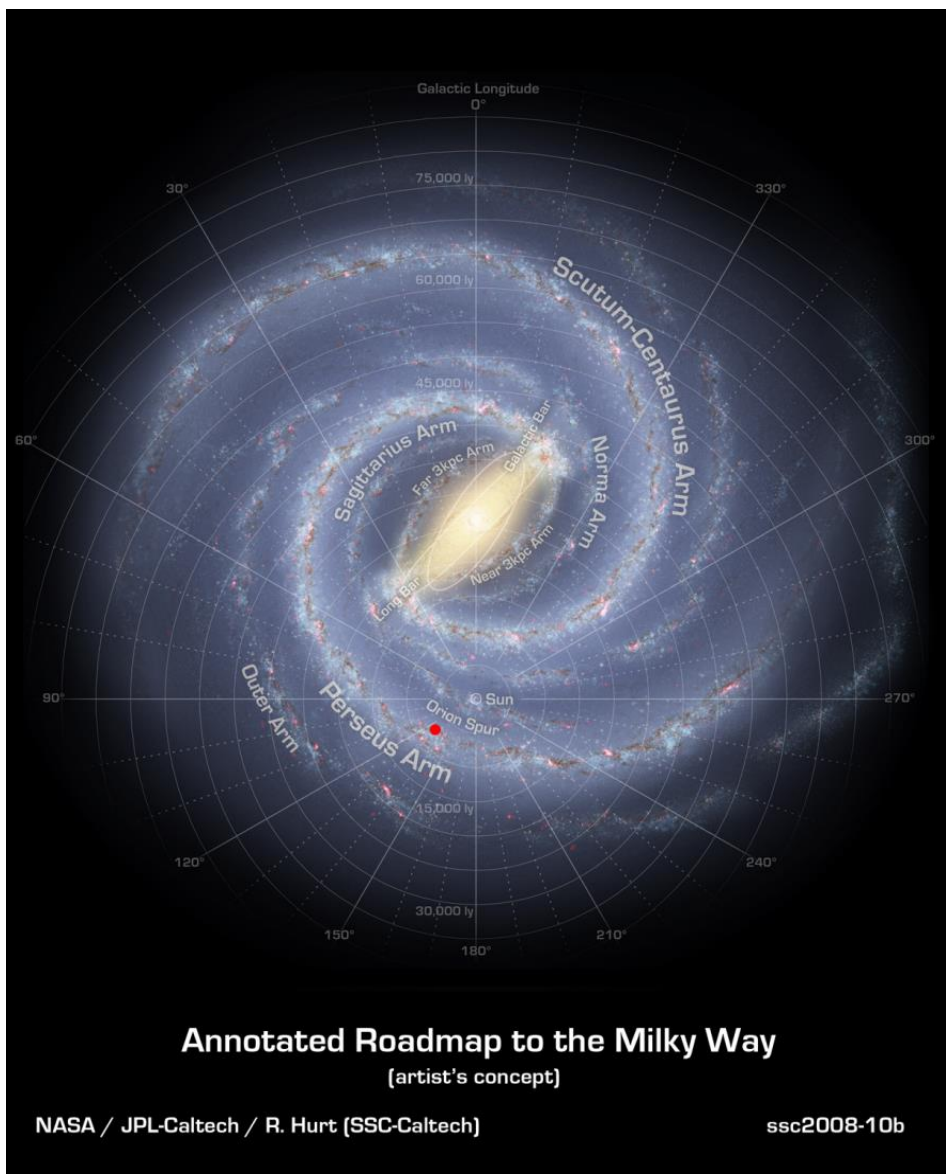

Figure 3. Annotated artist's concept illustrating the position of the major spiral arms of the Milky Way based on infrared images from NASA's Spitzer Space Telescope (21). The location of V909 Cas is marked with a red dot.

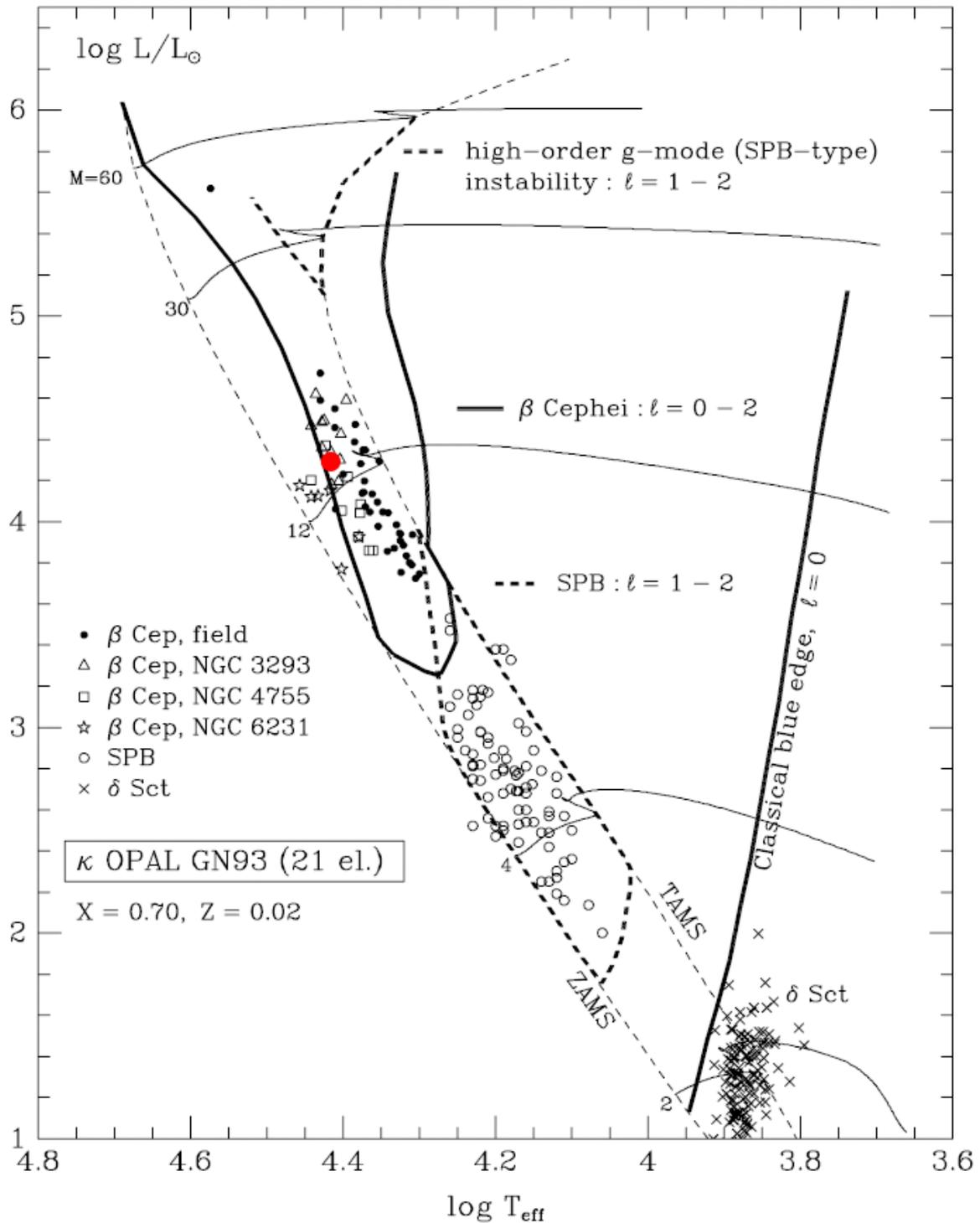

Figure 4. This shows the location of β Cep, slowly pulsating B (SPB) and delta Scuti stars in the upper main sequence of the H-R diagram based on evolutionary stellar models. The OB instability domain for β Cep stars is outlined in bold. ZAMS and TAMS are the Zero Age and Terminal Age Main Sequences where stars theoretically form and evolve away from the main sequence respectively. Some evolutionary trajectories for different stellar masses are shown. The location of V909 Cas within the distribution of known β Cep stars is shown in red. Diagram taken from Figure 3 in Pamyatnykh (2).